\documentclass[]{article}

\usepackage{frascatiphys}
\usepackage{graphicx}
\usepackage{amssymb}
\usepackage{amsmath}
\usepackage{amsfonts}
\usepackage{multirow}
\usepackage{hyperref}

\newcommand{\aap}{{A\&A}}

\newcommand{\apj}{{ApJ}}

\newcommand{\apjl}{{ApJ}}

\newcommand{\araa}{{ARAA}}

\newcommand{\mnras}{{MNRAS}}

\newcommand{\nat}{{Nat}}

\newcommand{\prl}{{Phys. Rev. L}}

\newcommand{\ssr}{{SSRv}}

\newcommand{\msun}{M$_{\odot}$}

\newcommand{\kms}{{km\,s$^{-1}$}}

\newcommand{{\ism}}{interstellar medium}
\newcommand{{\Snr}}{Supernova remnant}
\newcommand{{\snr}}{supernova remnant}
\newcommand{{\snrs}}{supernova remnants}
\newcommand{{\Snrs}}{Supernova remnants}
\newcommand{{\sn}}{supernova}
\newcommand{{\sne}}{supernovae}
\newcommand{{\pwn}}{pulsar wind nebula}
\newcommand{{\pwne}}{pulsar wind nebulae}
\newcommand{{\dsa}}{DSA}

\renewcommand{\deg}{{$^\circ$}}

\DeclareMathAlphabet\mathbfcal{OMS}{cmsy}{b}{n}

\def\idxg11{\index{G11.2-0.3}}

\def\idx79j{\index{SN 1979C}}

\begin{document}

\title{ 
WHAT SOURCES ARE THE DOMINANT GALACTIC COSMIC-RAY ACCELERATORS?
}
\author{
Jacco Vink\\
{\em 
Anton Pannekoek Institute \& GRAPPA, 
University of Amsterdam, Netherlands
} \\
}
\maketitle
\baselineskip=11.6pt
\begin{abstract}
Supernova remnants (SNRs) have long been considered to be the dominant source of Galactic cosmic rays, which implied that they provided most of the energy to power cosmic rays as well as being PeVatrons.
The lack of evidence for PeV cosmic rays in SNRs, as well as theoretical considerations, has made this scenario untenable. At the same time the latest LHAASO and other gamma-ray results suggest that PeVatrons lurk inside starforming regions. 

Here I will discuss why SNRs should still be considered the main sources of Galactic cosmic rays at least up to 10~TeV, but that the cosmic-ray data allow for a second component of cosmic rays with energies 
up to several PeV. This second component could be a subset of supernovae/SNRs, re-acceleration inside starforming regions, or pulsars. 
As a special case I show that the recent observations of Westerlund 1 by H.E.S.S. suggest a low value of the diffusion coefficient inside this region, which is, together with an Alfv\'en speed $\gtrsim 100$~\kms,
a prerequisite for making a starforming region  collectively  a PeVatron due to second order Fermi acceleration. 
\end{abstract}

\baselineskip=14pt

\section{Introduction}

The question about the origin of cosmic rays has been with us since their identification as an extraterrestrial source of ionization.\cite{hess12}
Cosmic rays have an energy distribution that is nearly a power law with index  $q\approx -2.7$  from 
$\sim 10^9$~eV to $\sim 10^{20}$~eV. The deviations in the spectrum from a pure, single power law---and compositional changes as a function of energy---provide information
on the the origin, transport and acceleration physics of cosmic rays.
Important features are the
steepening at  $\sim 3\times 10^{15}$~eV---the ``knee''---  and the flattening at $\sim 3\times 10^{18}$~eV---the ``ankle''. 
The  ``knee'' has long been thought to be the maximum energy that
protons can be accelerated to by the dominant class of Galactic cosmic-ray sources, whereas the ``ankle'' is  regarded to mark the transition from Galactic to extragalactic cosmic rays.
Although the sources of both Galactic and extragalactic cosmic rays remain topics of  debate, supernova remnants (SNRs) are commonly considered
to be the dominant source of Galactic, and active galactic nuclei (AGN) of extragalactic cosmic rays.

Here I will discuss that SNRs likely provide the bulk of Galactic cosmic rays, but that they do not accelerate up to the ``knee". Starforming regions provide observationally and theoretically
good conditions for acceleration of PeV cosmic rays, as  illustrated using new gamma-ray results  on Westerlund 1.

\section{Supernovae and the Galactic cosmic-ray energy budget}

SNRs are long known to be radio synchrotron sources, and over the last 30 yr many have  been identified as X-ray-synchrotron and very-high-energy (VHE) gamma-ray sources,  indicating the presence of  particles with energies of $10^8$~eV to $10^{14}$~eV. \cite{helder12a} 
The theory of diffusive shock acceleration\cite{malkov01} (DSA) provides the  theoretical framework to interpret the acceleration properties of SNRs.

For SNRs to be the dominant source class of Galactic cosmic rays, they must able to transfer a substantial fraction of their energy ($\sim 10^{51}$~erg) to cosmic rays. 
The Galactic energy density of cosmic rays is estimated to be $U_{\rm cr}\approx 1$~eV\,cm$^{-3}$, mostly concentrated around energies of $\sim 1$~GeV.\cite{webber98}
Composition measurements indicate that the typical
escape time of cosmic rays with these energies is $\tau_{\rm esc}\approx 1.5\times 10^7$~yr, whereas the Galactic diffusion coefficient is 
$D\approx 3\times 10^{28} (R/4~{\rm GV})^\delta $~cm$^2$s$^{-1}$, with $\delta\approx$0.3--0,7.\cite{strong07}
The one-dimensional diffusion length scale of cosmic rays is associated with the scale height of the cosmic-ray populations above and below the Galactic plane: $H_{\rm cr}=\sqrt{2D\tau_{\rm esc}}$.
The total energy budget to maintain the cosmic-ray energy density in the Galaxy  is then 
\begin{align}
\frac{dE_{\rm cr}}{dt}\approx &\frac{U_{\rm cr}\pi R^2_{\rm disc}2H_{\rm cr}}{\tau_{\rm esc} }
= U_{\rm cr} \pi R^2\sqrt{2D}\tau_{\rm esc}^{-1/2}\\\nonumber
\approx &1.0\cdot 10^{41}
\left(\frac{U_{\rm cr}}{{\rm eV\ cm^{-3}}}\right)
\left(\frac{R_{\rm disc}}{10~{\rm kpc}}\right)^2 \\\nonumber
&\  \ 
\left(\frac{D}{3\cdot 10^{28}~{\rm cm^2s^{-1}}}\right)^{1/2}
\left(\frac{\tau_{\rm esc}}{1.5\cdot 10^7~{\rm yr}}\right)^{-1/2}~{\rm erg\ s^{-1}},
\end{align}
with $R_{\rm disc}$ the typical radius of the Milky Way.

The supernova rate in the Milky Way is estimated to be 2--3 per century, providing $\dot{E}_{\rm sn}\approx 1.0\times 10^{41}$~erg~s$^{-1}$. 
So we find $\dot{E}_{\rm cr}\approx 10\% \dot{E}_{\rm sn}$.

\section{The PeVatron problem}
\label{sec:pevatronproblem}

SNRs as th\'e Galactic cosmic-ray sources are problematic when it comes to explaining  the cosmic-ray ``knee", which 
is often taken as evidence that the dominant cosmic-ray sources  should be able to accelerate protons beyond $10^{15}$~eV---i.e.t hey should be PeVatrons.  
However, SNRs are unlikely to be PeVatrons, both from an observational as well as from a theoretical point of view. 

Gamma-ray spectra of SNRs are reasonably well
described by power-law spectra with a break or a cutoff. The youngest known SNRs have gamma-ray spectrum  extending up to $\sim$10~TeV--100~TeV, but show a turnover in their spectra below 10~TeV.
This is both true for gamma-ray sources that are best modeled as hadronic gamma-ray sources---i.e. caused by pion production, such as Cas~A\cite{magic17_casa} and Tycho's SNR\cite{veritas17_tycho}---and for leptonic gamma-ray sources. 
Since typically the gamma-ray photon energy is  $\sim 10$\% of the energy of the primary particle, the gamma-ray spectra of young SNRs indicate that the cosmic-ray spectrum inside young SNRs cuts off below 100~TeV, well below the ``knee". 

Mature SNRs ($\sim 2000$--20,000~yr), such as W44 and IC443, have  breaks in their gamma-ray spectrum around 10-100 GeV, indicating the lack of cosmic-ray particles with
energies in excess of 1~TeV. In these  SNRs acceleration beyond 1~TeV has apparently stopped, and most previously accelerated particles have diffused out of the SNRs. 
Extended gamma-ray emission beyond the shock in the $\sim 2000$~yr old SNR RX J1713.7-3946 may indeed reveal escape of cosmic rays caught in the act.\cite{hess_rxj1713_escape}
An interesting, but peculiar, counter example is the relatively mature and luminous SNR N132D ($\sim 2500$~yr) in the Large Magellanic Cloud, for which no gamma-ray break or cutoff is detected below 
10~TeV.\cite{hess_n132d} This is in contrast to the much younger---but in many other ways comparable---SNR Cas A, which has a cutoff at $\sim 3$~TeV.\cite{magic17_casa}

The theoretical problems of acceleration of cosmic rays by SNRs beyond 1~PeV, using reasonable assumptions, are at least four decades old.\cite{lagage83,cristofari20} 
The  acceleration timescale according to DSA corresponds to the timescale for a shock-crossing cycle around the maximum energy:\cite{vinkbook}
\begin{equation}\label{eq:tau_acc}
\tau_{\rm acc}\approx \frac{8D_1}{V_{\rm s}^2},
\end{equation}
with $V_{\rm s}$ the shock velocity and $D_1$ the upstream diffusion coefficient. 
The diffusion coefficient for relativistic particles can be expressed as
\begin{equation}\label{eq:diffc}
D_1 = \frac{1}{3}\lambda_{\rm mfp} c=  \frac{1}{3}\eta r_{\rm g} c= \frac{1}{3}\eta \frac{E}{eB_1}c,
\end{equation}
with $c$ the particle speed/the speed of light, $r_{\rm g}$ the gyroradius, and $\lambda_{\rm mfp}=\eta r_{\rm g}$ a parametrisation of the mean-free path in terms of the
gyroradius. The smallest realistic value for  $D$ is for $\eta=1$, so-called Bohm diffusion, requiring a very turbulent magnetic field ($\delta B/B\approx 1$).

For SNRs we can  approximate $V_{\rm s}= m R_{\rm s}/t_{\rm snr}$, with $R_{\rm s}$ the shock radius and $t_{\rm snr}$ the SNR age. 
We can take Cas A as an example of a young cosmic-ray accelerator with $V_{\rm s}\approx 5500$~km/s\cite{vink22a},
$R_{\rm s}\approx 3$~pc,  $m=0.7$\cite{vink22a}, and  $B_1\approx 100~{\rm \mu G}$\cite{vink03a,voelk05}. 
For the acceleration timescale we write $\tau_{\rm acc}=ft_{\rm snr}$, with $f<1$ (typically $f=10$\%). 
Rewriting eq.~(\ref{eq:tau_acc}) gives:
\begin{align}\label{eq:emax}
E_{\rm cr,max} \approx &
\frac{3}{8}\eta^{-1} \frac{eB_1}{c} V_{\rm s}^2  \tau_{\rm acc} \approx
\frac{3}{8}\eta^{-1} \frac{eB_1}{c}m^2 f \frac{R_{\rm s}^2}{t_{\rm snr}}  \\\nonumber
= &1.4\times 10^{14}\eta^{-1}
\left( \frac{f}{10\%} \right)\left(\frac{m}{0.7}\right)^2 
\left(\frac{B_1}{100~{\rm \mu G}}\right)
\left( \frac{R_{\rm s}}{3~{\rm pc}}\right)^2
\left(\frac{t_{\rm snr}}{350~{\rm yr}}\right)^{-1}~{\rm eV}.
\end{align}
Optimistically taking $\eta=1$, we see that
Cas A cannot accelerate to PeV energies. The situation is better than theorized 30~yr ago\cite{lagage83}, 
because X-ray synchrotron filamentary widths in Cas A, as measured by the Chandra X-ray Observatory, provide evidence for amplified magnetic fields.\cite{vink03a,bell04}
 Moreover, X-ray synchrotron radiation by itself requires $\eta \approx 1$.\cite{aharonian99}

Athough two ingredients for large $E_{\rm cr,max}$---magnetic-field amplification and turbulence---appear to be present, they are not sufficient to make SNRs PeVatrons.
In fact, the optimism regarding $E_{\rm max}$ in SNRs is  tempered by the fact that 
 the measured gamma-ray cutoff energy for Cas A is  consistent with  $E_{\rm cr,max} \approx 10$~TeV,\cite{magic17_casa} rather than the expected $\sim$100~TeV. 
 This is peculiarly low for hadrons, as unlike electrons, they do not suffer radiative energy losses, but their maximum energy appears to be similar to the inferred maximum electron energy.\cite{vink03a}

\section{Alternative Galactic cosmic-ray source candidates}
\label{sec:alternatives}
Which energetic sources, other than SNRs, could be sources of Galactic cosmic rays?
Clearly, Galactic PeVatrons do exist as LHAASO recently has reported the detection of  PeV photons from various regions along the Galactic plane, including from the Crab Nebula or its pulsar.\cite{lhaaso21}.
Alternative source classes often discussed  
are pulsars\cite{fang12}, microquasars\cite{gallo05}, stellar winds\cite{webb85}, 
supernovae \cite{marcowith18}, superbubbles \cite{bykov92,parizot04,bykov20b}, and the supermassive black hole SGR A$^*$ \cite{hess_gc16}.

Most of these source classes are advocated based on the idea that they can accelerate particles up to the ``knee", but not all of them are capable of explaining the Galactic cosmic-ray energy density,
except ``supernovae" and ``superbubbles", which are grosso modo powered by the same source of energy as SNRs.
We ignore below SGR A$^*$, which may indeed be a PeVatron, and may have been more powerful in the past. However, the LHAASO results require the presence of PeVatrons throughout the Galactic plane, and not just in the Galactic center.

\subsection{Supernovae}
\label{sec:supernovae}

The supernova hypothesis usually assumes that a subset of supernovae, those exploding in a dense stellar wind, start accelerating almost immediately after the explosion.\cite{marcowith18} Typically these
are Type IIb and Type IIn supernovae, which comprise $\sim 10$\% of all supernovae. An important example of a potentially powerful accelerator was SN1993J, whose magnetic field at the shock
was estimated to be $\sim 10$~G \cite{fransson98} with an initial shock velocity of 20,000~\kms. 
Essentially the supernova hypothesis is  a ``very young SNR" hypothesis, as in supernovae such as SN1993J a bright SNR shell immediately develops in the dense wind of the progenitor. 

To get an idea of the maximum energy that can be reached under the right conditions, consider that the maximum distance traveled by 15,000 km/s shock is $R_{\rm s}=4.7\times10^{16}$~cm in one year, and if the wind velocity and densities are high Bell's instability\cite{bell04} could maintain a magnetic field of $\sim 1$~G near the shock. eq.~(\ref{eq:emax}) then gives $E_{\rm cr, max}\approx 1.3\times 10^{16}$~eV
reached within one year. Oservational proof for this hypothesis  would be the detection of VHE gamma rays from a radio-emitting supernovae, but so far only upper limits have been reported.\cite{hess_sne}

\subsection{Superbubbles and starforming regions}
\label{sec:superbubbles}
A substantial fraction of core-collapse supernovae probably explode inside of starforming regions.
Collectively these regions have, therefore, somewhat less supernova energy available than SNRs,
 but this is offset by the energy input provided by stellar winds (sect.~\ref{sec:winds}).
The recent LHAASO detection of PeV photons associated with starforming regions\cite{lhaaso21} provide observational evidence for the hypothesis that starforming regions/superbubbles
are PeVatrons.\cite{bykov20b} 
However, what needs to be proven is that the responsible multi-PeV cosmic rays  are not originating from the sources contained in starforming regions---supernovae, SNRs and stellar winds---but
that there are collective effects that keep on accelerating cosmic rays within the region as a whole. 
 In other words, are  starforming regions, from a cosmic-ray-acceleration point of view, more than the sum of their parts?

These ``collective" effects are in all likelihood due to second order Fermi acceleration \cite{fermi49}, which states that collisions of charged, relativistic cosmic rays with moving magnetic-field disturbances leads
to energy gains of
\begin{equation}\label{eq:dE_E}
\frac{\Delta E}{E}= \xi \left(\frac{v}{c}\right)^2,
\end{equation}
with $\xi\approx 1$ a parameter hiding the details of the interactions, and $c$ the speed of the relativistic particles. Since the magnetic disturbances are moving with the Alfve\'n speed we can set $v=V_{\rm A}$.
Second order Fermi acceleration takes into account gains due to head-on collisions, as
well as losses due head-tail collisions. 
Although second order Fermi acceleration is slower than DSA, acceleration in starforming regions can take up to millions of years, rather than the few thousand year timescale of SNRs.

There have been some calculations of the expected spectra of cosmic rays due to this mechanism, taking into account diffusion in phase-space.\cite{drury17}
Here I present a heuristic approach to obtain $E_{\rm max,cr}$. First note that the average ``collision time" is $\Delta t=\lambda_{\rm mfp}/c$, with $\lambda_{\rm mfp}$ the mean free path. In reality, there are no discrete collisions, but we use the same approach as
when we use a spatial diffusion coefficient, eq.~\ref{eq:diffc}. We can get rid of $\lambda_{\rm mfp}$ by stating $\Delta t=3D/c^2$, and 
the aforementioned  energy scaling  $D=D_0(E/E_0)^\delta$. 
Combing  $\Delta t$ with eq.~(\ref{eq:dE_E}) we obtain
\begin{equation}
\frac{1}{E}\frac{dE}{dt} =
  \xi \frac{c^2}{3D_0}\left(\frac{E}{E_0}\right)^{-\delta} \left(\frac{V_{\rm A}}{c}\right)^2,
\end{equation}
which for $0<\delta<1$ has the solution
\begin{equation}\label{eq:emax_2fermi}
E_{\rm max,cr}= E_1 + E_0\left[\frac{\xi c^2}{3D_0}  \left(\frac{V_{\rm A}}{c}\right)^2 \tau_{\rm acc} \right]^{1/\delta},
\end{equation}
with $\tau_{\rm acc}$ the timescale available for acceleration, and $E_1$ the injection energy of the particle. We can parameterize this for $\delta=1/2$, and $E_1\ll E_{\rm max,cr}$ as
\begin{equation}\label{eq:emax_2fermi}
E_{\rm max}\approx 1.4  \xi^2
 \left(\frac{V_{\rm A}}{150~{\rm km~s^{-1}}}\right)^4 \left(\frac{D(10~{\rm TeV})}{10^{26}~{\rm cm~s^{-1}}}\right)^{-2}\left(\frac{\tau_{\rm acc}}{1~{\rm Myr}}\right)^{2}~{\rm PeV}.
\end{equation}
Note that we need a large Alfv\'en velocity ($\sim 150$~\kms) and high level of magnetic-field turbulence---$D(10~{\rm TeV})\lesssim 10^{26}~{\rm cm^2s^{-1}}$---to create a PeVatron.\footnote{ 
Extrapolating the usual Galactic $D\approx 10^{28}~{\rm cm^2s^{-1}}$ at $\sim 1$~GeV to 10~TeV gives $D(10~{\rm TeV})\approx 10^{29}$--$10^{31}~{\rm cm^2s^{-1}}$,
at least three orders of magnitude larger than used in eq.~(\ref{eq:emax_2fermi}).}
Moreover, the particles need to be contained sufficient long to reach PeV energies.
I come back to this when discussing  Westerlund~1 (sect.~\ref{sec:wd1}).

\begin{figure}\label{fig:wind_energy}
\centerline{  \includegraphics[trim=70 70 100 100,clip=true,width=0.5\textwidth]{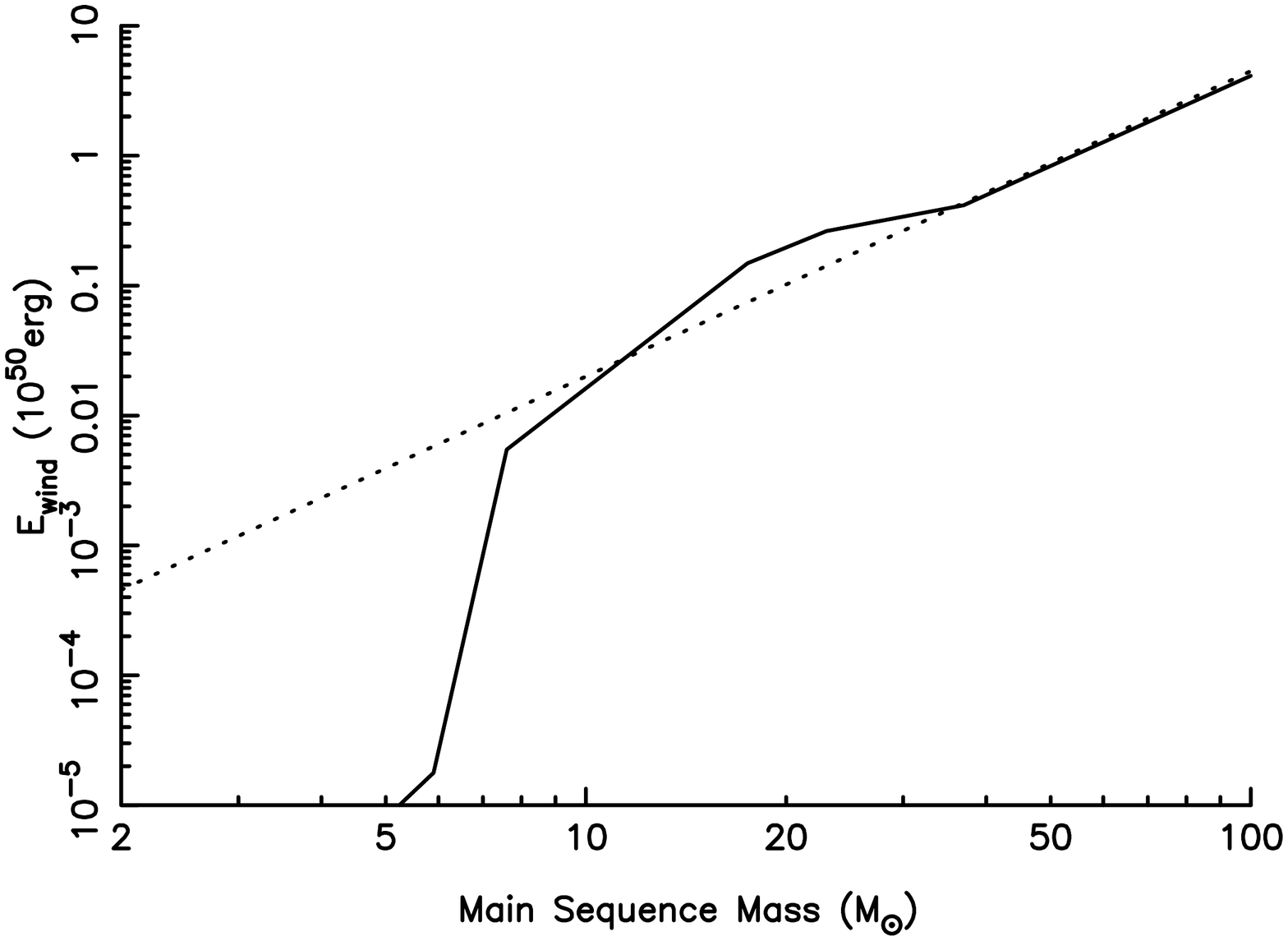}
}
  \caption[The wind energy of massive stars as a function of the main-sequence mass.]{The time-integrated, main-sequence wind energy of massive stars as a function of main-sequence mass.
    The dotted line indicates the scaling $E_\mathrm{wind}\propto M^{2.35}$. This figure is reproduced from\cite{vinkbook}.
  }
\end{figure}

\subsection{Stellar winds}
\label{sec:winds}

It is sometimes said that stellar winds may provide as  much kinetic energy as supernova explosion. The reality is somewhat more complicated. 

Best understood are the wind properties of massive main-sequence stars.\cite{vinkjs22} Fig.~\ref{fig:wind_energy} provides the time-integrated wind-energy of main-sequence stars.
Only for stellar masses approaching 100~\msun\ does the wind-energy approach the kinetic energy of  supernova explosions. But given the steepness of the initial mass function, the fraction of
these massive stars is very small. More common stars have $M_{\rm ms}\lesssim 25$~\msun, which provide $\lesssim 2\times 10^{49}$~erg. 

Things are more confusing beyond the main sequence. 
Most massive stars will become a red supergiant, or sometimes a yellow supergiant. These have enhanced mass losses, but their
wind velocities are low---$\sim 10$--50~\kms\ compared to 100~\kms\ up to $\sim 3000$~\kms\ for main-sequence winds---and hence provide little kinetic energy. However, some massive stars end their lives
in a Wolf-Rayet phase, characterized by large mass loss rates---$\dot{M}\sim 10^{-5}$~\msun$\ {\rm yr}^{-1}$---and very fast winds, 1000~kms--3000~\kms.  The most massive stars ($M_{\rm ZAMS}\gtrsim25$~\msun?) are likely ending in a Wolf-Rayet star
phase, but mass-stripping due to binary interactions  provides an alternative channel for  Wolf-Rayet star formation. 

So how much kinetic energy is associated with Wolf-Rayet stars? The measured wind velocities are in the range $v_{\rm WR}\approx $700~\kms--3000~\kms, with mass loss rates in the range
$\dot{M}_{\rm w}\approx$(0.5--6)$\times 10^{-5}~{\rm M_{\odot}~{yr}}^{-1}$.\cite{nugis00}
Typically a Wolf-Rayet star phase lasts a few 100,000 yr, implying a total time integrated energy of 
$E_{\rm w}=2\times 10^{50}(\dot{M}_{\rm w}/10^{-5}~{\rm M_\odot yr}^{-1})(v_{\rm w}/2000~{\rm km/s})^2(\tau_{\rm wr}/500,000~{\rm yr})$~erg. 
This is about $\sim 20$\% of the canonical supernova explosion energy. 
If we take Type Ibc supernovae to be explosions of Wolf-Rayet stars, we use theType Ibc supernova rate of 19\% of the overall rate\cite{li11}, to suggest that $\sim 20$\% of all massive stars become
Wolf-Rayet stars.
This implies that the Galactic power budget of Wolf-Rayet stars is  $\sim 4\%$ of the supernova power---small but not negligible. Moreover, 
in young (few Myr) stellar clusters like Westerlund 1 (see below), the power of Wolf-Rayet stars precedes the supernova power, as the most-massive stars
are in the Wolf-Rayet star phase, and the many less massive stars still need 5--20 Myr to evolve to the point of core collapse.

\subsection{Microquasars}
\label{sec:microquasars}

Microquasars are X-ray binaries containing an accreting neutron star or black hole, that develop jet outflows during certain accretion phases. They are regarded as nearby analogues to
the radio galaxies and quasars (i.e. AGN). 
Given that AGN are the most likely sources for extragalactic cosmic rays, detected with energies up to $\sim 10^{20}$~eV, it is not unreasonable to assume that also microquasars
are good Galactic accelerators. Indeed, radio and gamma-ray emission shows that they do accelerate at least electrons.\cite{hess05_ls5039} 
The shell found around Cygnus X-1 has also been used
as evidence for energetic jets from these systems.\cite{gallo05}
However, microquasars are a much less abundant gamma-ray source class than SNRs and pulsar-wind nebulae.\cite{hess18_gpsurvey}
Moreover, the number of systems available at any given moment seem to be 50--100.\cite{dubus17} Together with a typical jet power of $\sim10^{38}$~erg/s,
this implies a  typical Galactic kinetic power to be attributed to  microquasars of $\dot{E}_{\rm \mu q}\approx 10^{40}$~erg/s. If 10\% of that power is transferred to cosmic rays, microquasars fall a factor hundred short of
maintaining the Galactic cosmic-ray energy density.

\subsection{Pulsars}
\label{sec:pulsars}

Pulsar wind nebulae are among the most common Galactic gamma-ray sources. However, the canonical theory is that pulsar wind nebulae (PWNe) contain mostly electrons/positrons created by pair
creation in the pulsar magnetospheres. Clearly they are efficiently accelerating electrons/positrons, and the archetypal Crab pulsar/PWN is even a confirmed PeVatron, given the detection of PeV photons from this source by LHAASO.\cite{lhaaso21}
It is possible that the pulsar winds do not solely consists of electrons/positrons and Poynting flux, but may also contain hadrons \cite{amato14},
potentially making pulsars hadronic PeVatrons.
However, they are unlikely to be the dominant source of Galactic cosmic rays from an energy-budget point of view.

I illustrate this by pointing out that the pulsar birth rate is similar, but somewhat smaller than the supernova rate, i.e. about $\sim 2$ per century. 
For normal pulsars the energetic output comes at the expense of the rotational energy of the neutron star. 
The total initial rotational energy available is 
\begin{equation}
E_{\rm rot} = \frac{1}{2} I\Omega_0^2=\frac{2\pi^2}{P_0^2}I\approx 8\times 10^{48} \left(\frac{I}{10^{45}~{\rm g~cm^2}} \right)\left(\frac{P_0}{50~{\rm ms}}\right)^{-2}~{\rm erg},
\end{equation}
with $P_0=2\pi/\Omega_0$ the initial spin period and $I$ 
the moment of inertia.
For pulsars to compete energetically with supernovae, they need $P_0\lesssim 5$~ms. However, the initial spin period inferred from population synthesis
models indicate a  much longer initial period, 50--100~ms, or even 300~ms.\cite{faucher06} Taking $P_0\gtrsim 50$~ms gives
$\dot{E}_{\rm psrs}\lesssim 5\times 10^{39}$~erg/s, insufficient for powering Galactic cosmic rays.

The above considerations suggest that pulsars are not prominent sources of Galactic cosmic rays, even if they accelerate hadrons. But in the latter case they could  be a source of PeV protons.
Pulsars are likely an important, perhaps even dominant, contributor to the electron-/positron cosmic-ray population. Moreover, the pulsar wind nebulae (PWNe) and pulsar wind haloes \cite{linden22}
constitute an important class of VHE gamma-ray sources.\cite{hess18_gpsurvey} There is no contradiction between being prominent gamma-ray sources and not providing enough energy to power Galactic cosmic rays: energetic electrons/positrons are radiatively much more efficient than energetic protons. 

\section{Do the sources of Galactic cosmic rays need to be PeVatrons?}

For a long time it was considered that the dominant sources of Galactic cosmic rays must fulfill both  the cosmic-ray energy budget and be PeVatrons.
The reason was a lack of cosmic-ray spectral features between $\sim 1$ GeV and $3\times 10^{15}$~eV, 
whereas if there were two or more source classes fulfilling together these criteria, we would expect  some breaks in the cosmic-ray spectrum. 

It has now time to reconsider that idea, because of evidence that the cosmic-ray spectrum below the ``knee" is  not so featureless. 
First of all, the proton cosmic-ray spectrum has a different slope than the helium cosmic-ray spectrum.\cite{adriani11}
This suggest two different origins, although both could originate from SNRs in different environments, or forward shock versus reverse shock acceleration.\cite{ptuskin13}
Secondly, the latest cosmic-ray measurements\cite{ams15,dampe19,kobayashi21} indicate that the proton cosmic-ray spectrum hardens around $\sim 0.7$~TeV and softens again around $15$~TeV.\cite{lipari20} The situation regarding the proton spectrum around the ``knee" is not clear. Certainly the break around $15$~TeV is consistent with the maximum energy young SNRs can accelerate
protons to. 

These results open up the possibility that SNRs indeed provide the bulk of Galactic cosmic rays, a scenario that agrees with the cosmic-ray energy budget, and that other
source classes---or subclasses of SNRs, including supernovae---are responsible for cosmic rays from 15 TeV up to the ``knee". Note that this requires PeVatrons to have  harder (flatter slope)
spectra than SNRs, in order for these PeVatron sources to be subdominant around 1~GeV.

\section{Observational candidate PeVatrons: starforming regions}

Since the detection of PeV photons from Galactic plane sources by LHAASO\cite{lhaaso21} and some associations with starforming regions,
starforming regions  demand more attention as sources of cosmic rays and  being potentially PeVatrons
themselves (sect.~\ref{sec:superbubbles})---as opposed to merely containing PeVatrons.

A notable source of PeV photons  is 
 LHAASO J2032+4102, which is positionally associated with the Cygnus Cocoon surrounding the Cyg OB2 star cluster. 
Imaging atmospheric Cherenkov telescope arrays also provide evidence for the existence of  PeVatrons associated with starforming regions, and provide a better
angular resolution. 
For example, recently H.E.S.S. detected VHE gamma-ray emission from HESS J1702-420(A) up to energies of 100~TeV, implying primary particles of up to or beyond 1~PeV.\cite{hess21_1702}
But the nature of this source, which has a complex gamma-ray morphology, is not entirely clear.  
Although not a confirmed PeVatron, another interesting starforming region is Westerlund 1/HESS J1646-458, for which the gamma-ray spectrum extends at least up to 50~TeV.\cite{hess22_wd1}

\begin{figure}
\centerline{
\includegraphics[trim=0 -20 0 0,clip=true,width=0.45\textwidth]{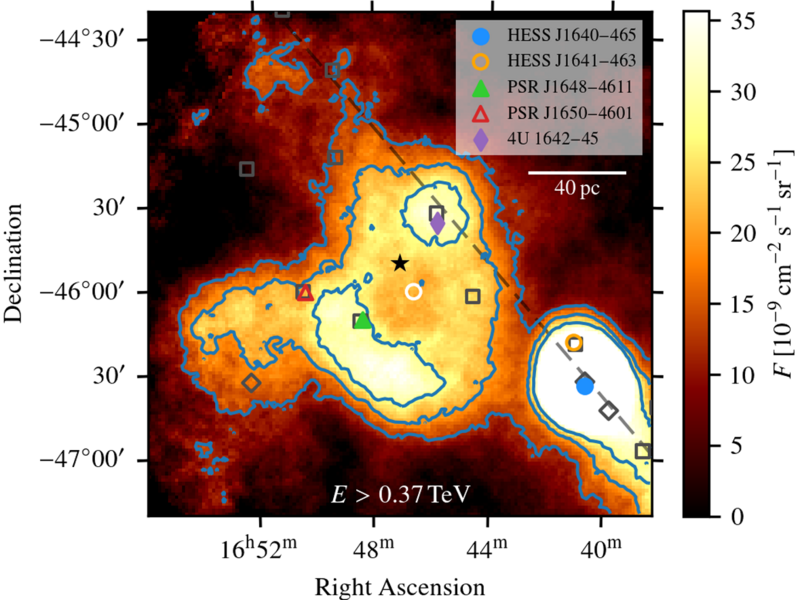}
\includegraphics[trim=50 50 50 50,clip=true,width=0.55\textwidth]{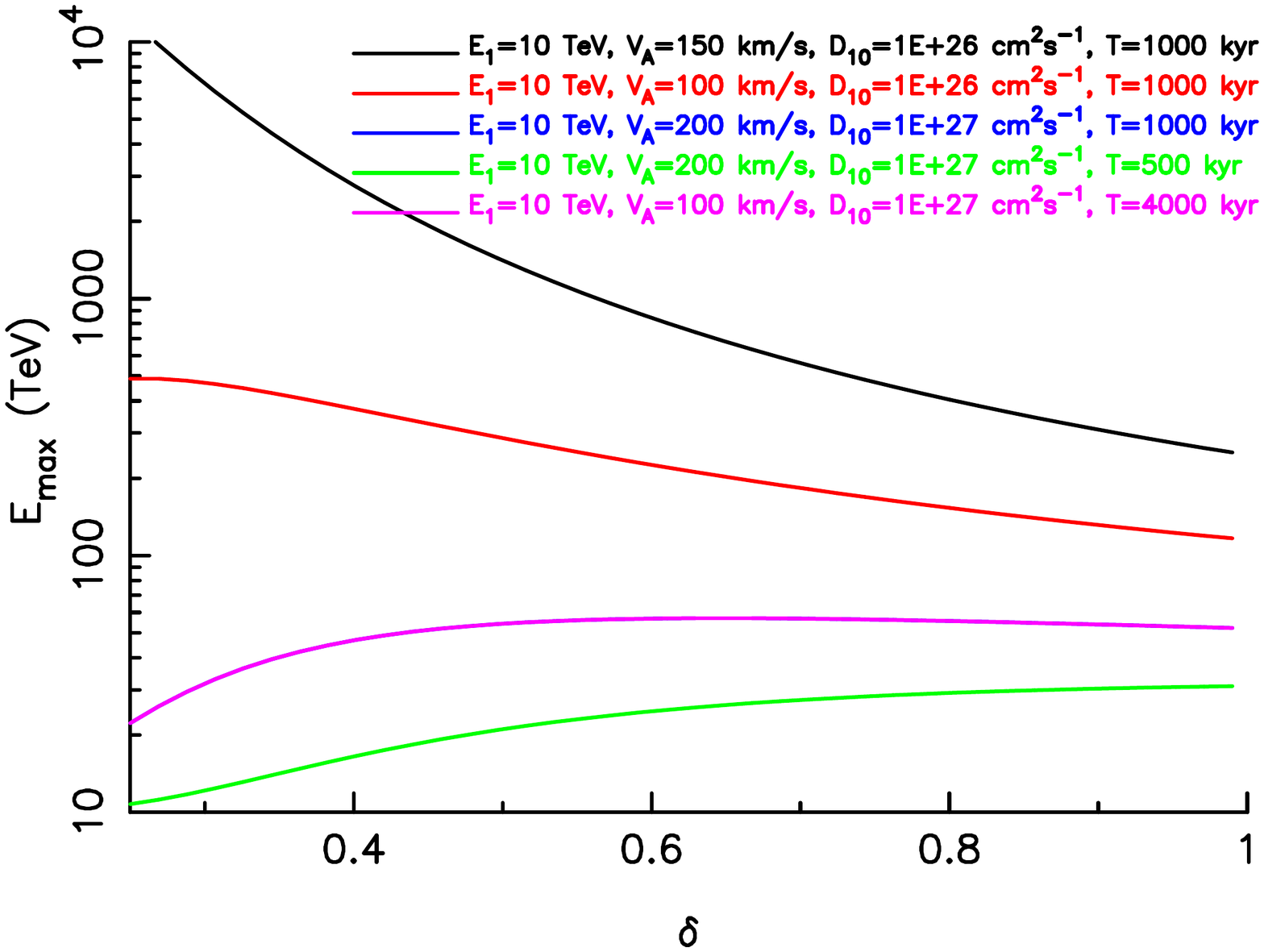}
}
\caption[]{
Left: VHE gamma-ray map of Westerlund 1 by H.E.S.S. \cite{hess22_wd1}
The cluster postition is indicated by a black star.
  (Credit: \href{https://www.mpi-hd.mpg.de/hfm/HESS/pages/home/som/2022/08/}{H.E.S.S. Collaboration})
Right: Maximum cosmic-ray energy by 2nd order Fermi acceleration according to eq.~(\ref{eq:emax_2fermi}) as a function  of $\delta$ and for various
values of $D(10~{\rm TeV})$, $V_{\rm A}$  and $\tau$.\label{fig:wd1}
}
\end{figure}

\subsection{Very-high energy gamma-ray emission from Westerlund 1}
\label{sec:wd1}
Westerlund 1 is the most massive young massive star cluster known in the Milky Way, $\sim 10^5$~\msun, which hosts many Wolf-Rayet stars and other evolved massive stars. Its age has been
estimated to be $\sim 4$~Myr, but recent work suggests an age spread up to 10~Myr.\footnote{See the recent H.E.S.S. paper for references.\cite{hess22_wd1}}
The VHE gamma-ray emission does not originate from the stellar cluster itself, but  from a large shell-like region
surrounding it, with peak radius of $\sim 0.5$\deg (fig.~\ref{fig:wd1}). At a distance of 4~kpc the shell has a physical radius of $\sim 35$~pc, whilst Westerlund 1 itself is much more compact, $\sim 1$~pc.

In the H.E.S.S. publication an interpretation is favored in which the shell-like emission is associated with a collective-wind termination shock,\cite{hess22_wd1,morlino21} 
as a physical shell, consisting of swept-up gas, was considered unlikely, because the gamma-ray structure is too
small for the energy input from Wolf-Rayet-star winds ($\dot{E}_{\rm WR}\gtrsim 10^{39}~{\rm erg~s^{-1}}$ \cite{muno06b}) and the age of the cluster,
as based on the stellar-bubble expansion model by Weaver et al.\cite{weaver77}.

However, some of the assumptions used  to calculate the putative shell size may not be correct.  
 The cluster age of $\sim 4$~Myr is a poor indicator for the total energy and shell creation timescale available for making a shell. 
As noted in sect.~\ref{sec:winds}, the Wolf-Rayet star phase
typically lasts for a few 100,000 yr, consistent with being about 2--10\% of the stellar life time.  Taking $\tau_{\rm SB}\approx 200,000$~yr  
reduces the radius of the collective wind bubble from
$R_{\rm SB}=185 (n_{\rm H}/5)^{-1/5}$~pc to $R_{\rm SB}=31 (n_{\rm H}/5~{\rm cm^{-3}})^{-1/5}$~pc. There is another reason to be suspicious about the predicted size of the bubble: observationally superbubbles appear to be smaller than predicted by  wind-bubble expansion theories. This may be 
due to internal dissipation and radiative losses, as well as back pressure from the ambient medium.\cite{krause14b,chevalier99}

The lack of an association of the VHE gamma-ray shell-like structure with an HI or CO structure is intriguing, but should also not be taken as evidence against the presence of a physical shell, as the intense UV light from the
OB and Wolf-Rayet stars in Westerlund 1 likely photo-dissociated/ionize a large part of the surrounding CO and neutral hydrogen.

Assuming that Westerlund 1 itself is indeed accelerating cosmic rays, or further accelerating particles pre-accelerated by the colliding winds and past SNRs, what can we learn from the VHE gamma-ray spectrum
and morphology measured by H.E.S.S.?

The fact that the VHE gamma-ray morphology seems to have no or little dependence on gamma-ray energy suggest that the cosmic-ray particles have been well mixed within the emitting
region and may have been accelerate throughout the bubble.
The best-fit cutoff energy of $\approx 40$~TeV---corresponding to proton energies above 100 TeV--indicates that escape is only important for particles above $\sim 100$~TeV.
Equating the radius of the shell to the threedimensional diffusion length scale, $R_{\rm shell}=\sqrt{6D\tau}$, and using $\tau\approx 200,000$~yr, we can estimate the diffusion coefficient 
at $\sim 100$~TeV to be $D\approx R_{\rm shell}^2/(6\tau)\approx 3\times 10^{26}(\tau/200,000~{\rm yr})^{-1}~{\rm cm^2s^{-1}}$, corresponding to $D\approx 10^{26}~{\rm cm^2s^{-1}}$ at 10 TeV. 
This is  close to the value for which considerable energy gain
due to second order Fermi acceleration is expected (eq.~\ref{eq:emax_2fermi})!

The maximum energy as a function of the energy dependence of the diffusion coefficient, i.e. $\delta$, is shown in fig.~\ref{fig:wd1} (right) for different valued of $D$ and $V_{\rm A}$. 
It is clear that $V_{\rm A}\gtrsim 100$~\kms. From $V_{\rm A}=B/\sqrt{4\pi \cdot1.4\cdot n_{\rm H}m_{\rm p}}$ we find that we need a low internal density:
$n_{\rm H} \lesssim 0.2 (B/10~{\rm \mu G}) (V_{\rm A}/100~{\rm km~s^{-1}})^{-1}~{\rm cm^{-3}}$.  However, even  densities of $10^{-3}~{\rm cm^{-3}}$ are possible in superbubbles of a few Myr old.\cite{krause14a}
Moreover,  $10~{\rm \mu G}$  is rather modest and corresponds to an internal magnetic-field energy density within the shell of $2\times 10^{49}$~erg.
Interestingly, using eq.~\ref{eq:diffc} with $B\approx 10~{\rm \mu G}$  and $D(10~{\rm TeV})=10^{26}~{\rm cm^2s^{-1}}$ provides a value of $\eta\approx 3$, which is very close to Bohm diffusion.
Stronger magnetic fields are still consistent with the energy budget, required Alfv\'en speeds and a PeVatron hypothesis. Much lower strengths lead to inconsistencies, such as $\eta <1$, or $D\gg 10^{26}~{\rm cm^2s^{-1}}$.

To summarize, the VHE gamma-ray spectra and morphology of Westerlund 1 imply a small diffusion coefficient, whereas a  low density inside the shell and amplified, turbulent magnetic field likely result
in fast Alfv\'en velocity, setting the right conditions to (re)accelerate  cosmic rays injected by primary sources to well beyond $100$~TeV.
As such Westerlund 1 may provide a model for other starforming  regions as PeVatrons.

\section{Conclusion}

I have argued here that SNRs likely are the dominant sources of Galactic cosmic rays below $\sim 10$~TeV, as
both observational and theoretical results are consistent with young SNRs being able to accelerate to at least  this energy.
Moreover,
the latest
cosmic-ray measurements indicate that there is is a hardening of the proton cosmic-ray spectrum around  $\sim 10$~TeV, indicating the presence of  additional sources of Galactic cosmic rays,
which may be responsible for cosmic rays up to the ``knee".

If these additional sources accelerate cosmic rays with a relatively hard spectrum, this source class does not violate the energetic constraints.  Like SNRs, this PeVatron source class  may rely on the power input of supernovae,
be its a subclass of SNRs, core-collapse supernovae exploding inside the dense stellar wind of the progenitor star, or the collective power of supernovae and stellar winds in starforming regions/superbubbles.
Energetic pulsars should not be discarded as hadronic PeVatrons, but it first remains to be proven that pulsars are hadronic- and not just leptonic-accelerators.
Both pulsars and  starforming regions/superbubbles as PeVatron sources are consistent with the latest detections of Galactic PeV gamma-ray  sources by LHAASO.

Several starforming regions/superbubbles have been associated with PeVatron candidate sources. It is not yet clear whether these are PeVatrons by themselves,
or whether they merely contain(ed) PeVatrons sources (in the past), such as the aforementioned subclass of SNRs/supernovae and hadronic pulsar accelerators.

I argue that the recently  reported VHE gamma-ray properties of HESS J1646-458---associated with Westerlund 1---indicates a small  internal diffusion coefficient; small enough to accelerate protons up to the ``knee" in a few 100,000 yr, provided that the Alfv\'en speed is $\gtrsim 100$~\kms.
This suggests that starforming regions/superbubbles could be themselves PeVatrons. As cosmic-rays source,  starforming regions may  be more than the sum of their parts.

\vskip 1.2em
{\noindent\bf Acknowledgements}\\
I am indebted to my involvement in and membership of  H.E.S.S. However, this text is written on  personal title.
 The author is supported 
 by funding from the European Union's Horizon 2020 research and innovation program, grant agreement No.101004131 (SHARP).

\end{document}